%% file: ms_rev3.tex
\documentclass[preprint]{aastex61}

\accepted{25 January  2018}

\submitjournal{ApJ}

\begin{document}

\title{Resolving the Inner Arcsecond of the  RY Tau Jet with {\em HST}}

\correspondingauthor{Stephen L. Skinner}
\email{stephen.skinner@colorado.edu} 

\author{Stephen L. Skinner}
\affiliation{Center for Astrophysics and 
Space Astronomy (CASA), Univ. of Colorado,
Boulder, CO, USA 80309-0389}

\author{P. Christian Schneider}
\affiliation{Hamburger Sternwarte, Universit\"{a}t Hamburg, Gojenbergsweg 112,
21029 Hamburg, Germany}

\author{Marc Audard}
\affiliation{Dept. of Astronomy, 
University of Geneva, Ch. d'Ecogia 16, CH-1290 Versoix, Switzerland}

\author{Manuel  G\"{u}del}
\affiliation{Dept. of Astrophysics, Univ. of Vienna, 
T\"{u}rkenschanzstr. 17,  A-1180 Vienna, Austria} 

%
\newcommand{\ltsimeq}{\raisebox{-0.6ex}{$\,\stackrel{\raisebox{-.2ex}%
{$\textstyle<$}}{\sim}\,$}}
%
\newcommand{\gtsimeq}{\raisebox{-0.6ex}{$\,\stackrel{\raisebox{-.2ex}%
{$\textstyle>$}}{\sim}\,$}}
%
\begin{abstract}
\small{
Faint X-ray emission from hot plasma (T$_{x}$ $>$ 10$^{6}$ K) has been detected
extending  outward a few arcseconds along the optically-delineated jets of some 
classical T Tauri stars including RY Tau. The mechanism and location where the 
jet is heated to X-ray temperatures is unknown.  We present high spatial resolution
{\em Hubble Space Telescope } ({\em HST}) far-ultraviolet long-slit observations of  
RY Tau with the slit aligned  along the jet.  The primary objective was to 
search for C IV emission from warm plasma at T$_{CIV}$  $\sim$ 10$^{5}$ K
within the  inner jet ($<$1$''$) that cannot be fully-resolved  by  X-ray telescopes.
Spatially-resolved C IV emission is detected in the blueshifted jet extending 
outward from the star to 1$''$  and in the redshifted jet out to 0$''$.5. 
C IV line centroid shifts give a radial  velocity in the blueshifted jet  
of $-$136 $\pm$ 10 km s$^{-1}$ at an offset of 0$''$.29 (39 au) and 
deceleration outward is detected. The deprojected jet speed is subject  
to uncertainties in the jet inclination but values $\gtsimeq$200 km s$^{-1}$ are likely.
The  mass-loss rate in the blueshifted jet is at 
least $\dot{M}_{jet,blue}$ = 2.3 $\times$ 10$^{-9}$  M$_{\odot}$ yr$^{-1}$, 
consistent with  optical determinations. We use the {\em HST} data along with
optically-determined jet morphology to place meaningful constraints on
candidate jet-heating models including a hot-launch model in which
the jet is heated near the base to X-ray temperatures by an unspecified
(but probably magnetic) process, and downstream 
heating from shocks or a putative jet magnetic field.
}
\end{abstract}


\keywords{ISM: jets and outflows --- stars: individual (RY Tau)  --- stars: mass loss ---  
stars: pre-main sequence  --- ultraviolet: stars}


\section{Introduction}
Highly-collimated outflows or ``jets'' are a widespread
phenomenon in astrophysics and have been detected in
a diverse range of objects including young stars and protostars, 
compact binaries, planetary  nebulae, and AGNs (reviewed by Livio 1999). 
In star-forming regions, jets are usually associated with accreting protostars,
classical T Tauri stars (cTTS), and Herbig Ae/Be stars and are
revealed in the optical as shock-excited Herbig-Haro (HH)
objects. As such, jets provide an excellent laboratory
for studying shock physics. Jets from young stars
inject turbulence into the surrounding molecular cloud.
Jet rotation, along with other mechanisms, may remove
angular momentum from the accretion disk (reviewed by Frank et al. 2014).
Possible detections of jet rotation in a few young stellar objects
have been reported, a recent example being the 
protostellar system HH 212 (Lee et al. 2017).
But searches for jet rotation signatures in RY Tau,
the cTTS of primary interest here, have so far not yielded
a high-confidence ($>$3$\sigma$) detection (Coffey et al. 2015).

Jets from young stars have traditionally been identified and
studied  using optical and radio telescopes, both of which can
provide sub-arcsecond resolution. Optical studies of  forbidden 
spectral lines yield information on properties of cool (T $\sim$ 10$^{4}$ K)
plasma in shocked jets. Somewhat surprisingly, much hotter X-ray plasma at 
temperatures  T$_{x}$  $\gtsimeq$ 10$^{6}$ K has now been detected at small offsets 
($\leq$4$''$) along the jets of some young stars such as  DG Tau (G\"{u}del et al. 2005, 2008), 
Z CMa (Stelzer et al. 2009),  RY Tau (Skinner, Audard, \& G\"{u}del 2011),
and possibly RW Aur A (Skinner \& G\"{u}del 2014). 
Faint extended jet emission has also been reported in the vicinity of a few
jet-driving protostars such as  
L1551 IRS 5 (Bally, Feigelson, \& Reipurth 2003;
Schneider, G\"{u}nther, \& Schmitt 2011)
and Herbig-Haro objects (Pravdo et al. 2001, 2004; Favata et al. 2002;
Grosso et al. 2006).

The mechanism(s) by which jets from young stars are launched,
collimated, and heated are not fully understood but it is 
thought that magnetic fields play a key role. Shocks undoubtedly
also play a role in jet heating but it is not yet clear that shocks
alone can heat jets from young stars to X-ray emitting temperatures.
In order to reach X-ray temperatures of a few MK, jet speeds 
(and shock speeds) of at least v$_{s}$ $\approx$ 300 - 400 km s$^{-1}$ are required.
But some jet-driving TTS for which faint extended X-ray jet
emission has been detected apparently do not achieve such high
jet speeds and other heating mechanisms besides shocks may be at work.

Of specific interest here is RY Tau, a relatively massive cTTS whose
properties are summarized in Table 1. It is remarkable in several
respects, being a  rapid rotator with  $v$sin$i$  = 52 $\pm$ 2 km s$^{-1}$ 
(Petrov et al. 1999) and a bright variable  X-ray source indicative
of strong magnetic activity (Skinner, Audard, \& G\"{u}del 2016). 
It is undergoing disk accretion 
(Schegerer et al. 2008; Agra-Amboage et al. 2009) and mass loss is present 
from a wind (Kuhi 1964; G\'{o}mez de Castro \& Verdugo 2007) and a bipolar jet known as
HH 938 (St.-Onge \& Bastien 2008). Searches for a companion
have so far yielded negative results (Leinert et al. 1993; Schegerer et al. 2008;
Pott et al. 2010).
 
Continuum-subtracted H$\alpha$ images obtained by St. Onge \& Bastien (2008)
traced the approaching (blueshifted) jet  outward to 31$''$ from the star  along
P.A. $\approx$295$^{\circ}$ and inward to within 1$''$.5 of the star.
These images also show more distant features attributed to the receding  counterjet  
at offsets of $\approx$2$'$.8 - 3$'$.5 from the star.
A sub-arcsecond resolution  ground-based study of the jet using the 
Canada-France-Hawaii Telescope by Agra-Amboage et al. (2009) revealed
high-velocity ($\gtsimeq$100 km s$^{-1}$) jet plasma extending outward
to at least 1$''$ - 2$''$ from the star. After taking into account projection
effects due to the jet inclination angle to the line-of-sight  
($i_{jet}$ = 61$^{\circ}$ $\pm$ 16$^{\circ}$)
they estimated  the most probable deprojected jet speed to be 
v$_{jet}$ $\approx$ 165 km s$^{-1}$. This value is too low to explain the
faint  extended X-ray emission seen in deconvolved {\em Chandra} images
overlapping the blueshifted jet in terms of a jet that is heated only
by shocks (Skinner et al. 2011).

The above results raise the question of whether hotter and higher speed jet plasma
could have escaped detection in ground-based optical studies. We investigate
this possibility here by acquiring spatially-resolved far-ultraviolet (FUV) 
spectroscopic images of the inner jet using the {\em HST} Space Telescope
Imaging Spectrograph (STIS). 
As discussed in more detail below, the new STIS spectra show that the jet 
speed is higher than measured in  optical studies  but is still only
marginally sufficient at best to produce  shock-induced X-rays. 

\input{table1.tex}


\section{Observations and Data Reduction}

The {\em HST} STIS long-slit observations of RY Tau were obtained 
in December 2014  using  the Multi-Anode Micro-channel Array (MAMA) 
detectors. In contrast to previous STIS observations (Calvet et al. 2004),
the  52$''$ $\times$ 0.$''$2  slit was aligned along the optical 
jet axis in order to sample emission along the jet.
Table 2 summarizes the observations and basic
instrument properties. 

We focus here on the medium-resolution G140M spectrum which 
spectrally-resolves the  C IV resonance doublet whose 
reference wavelengths are listed in the {\em CHIANTI} atomic
database (Del Zanna et al. 2015) as $\lambda_{lab}$ =
1548.189/1550.775~\AA\footnote{Accessible in electronic form at:~
http://www.chiantidatabase.org/chianti\_linelist.html}.
The G140M grating provides detailed information on line centroid
wavelengths (velocities) and shapes with a full-width half-maximum (FWHM)
spectral resolution of $\approx$0.075~\AA , equivalent to $\approx$15 km s$^{-1}$ 
at 1550~\AA. The C IV doublet  traces warm plasma at a characteristic 
temperature T$_{CIV}$ $\sim$ 10$^{5}$ K and has been brightly-detected in 
previous {\em HST} observations of RY Tau (Calvet et al. 2004; Ardila et al. 2013).

Data were analyzed using PyRaf v. 2.1.6. Flat-fielded calibrated 
two-dimensional (2D) image science files (*flt.fits) for each 
exposure were provided by the Space Telescope Science Institute (STScI)
production pipeline.
The flat-fielded images for exposures taken at dithered offset 
positions were shifted and aligned with images for undithered 
exposures. The set of aligned flat-fielded images was then
combined into a total exposure 2D spectral image (Fig. 1) using 
the  Pyraf task {\em x2d}. Calibrated and background-subtracted
one-dimensional (1D)  spectra giving flux density as a function of
wavelength  were extracted from flat-fielded calibrated images using the 
task {\em x1d}. Wavelengths in the 2D spectral image
and 1D spectra are in the heliocentric reference frame.
We extracted spectra centered on the stellar trace
for individual exposures (typical exposure time 815 s) to check
for time-variable C IV emission and for the total usable  summed exposure
of 10,765 s. In addition, 1D spectra based on the total
exposure were extracted at incremental offsets 0$''$.145 (5 MAMA pixels) 
along the  blueshifted and redshifted jets. 
The spectra at each offset were extracted using a spatial
binsize of 5 MAMA pixels (full-width). This strategy provides 
complete non-overlapping spatial sampling along the jet.
The choice of a 5-pixel step-size  and a 5-pixel 
spatial binsize for each spectral extraction is a compromise
between the need to obtain good spatial sampling while
capturing sufficient counts in each spectrum to reliably
measure line parameters. Taken together, the 1D spectra
provide spatially-resolved information on the jet radial velocity, 
line width, and flux as a function of projected distance from the star.  
Decreasing signal-to-noise (S/N) ratio  in the C IV lines
toward larger offsets restricted our 1D spectral analysis to offsets of
$\leq$1$''$ in the blueshifted jet and $\leq$0$''$.5 in the 
redshifted jet. 

\input{table2.tex}

\section{Results}

\subsection{STIS G140M 2D Spectral Image}
The STIS G140M 2D total-exposure spectral image at wavelengths
near C IV  is shown in Figure 1. 
The C IV doublet lines are clearly detected and spectrally-resolved.
Spatial extension is visible in both doublet lines along the blue
and redshifted jets. The blueshifted (approaching) jet is traced out to 
$\approx$1$''$ ($\approx$134 au)  from the star and the redshifted 
jet out to $\approx$0$''$.5 ($\approx$67 au).
The maximum intensity (peak pixel) of both lines is associated
with the  brighter blueshifted jet lobe. The spatial position of the peak 
pixel in both doublet lines is offset slightly from the stellar
trace in the direction of the blueshifted jet. 
The offset in the
1547 - 1551 \AA~ range is 0.3 $\pm$ 0.5 MAMA pixels which is 
of low significance compared to the $\pm$0.2 pixel uncertainty in 
the location of the stellar trace on the detector as determined from 
calibration image fiducial bars.

As discussed further below, the C IV lines are quite broad at
the stellar position but become narrower and more symmetric
at increasing offsets along the jet. Several slightly-blueshifted 
moderately-broadened fluorescent 
H$_{2}$ lines are also detected, the
brightest of which is R(11) 2-8 ($\lambda_{lab}$ = 1555.89 \AA).
This line shows faint spatial extension out to at least $\approx$5 MAMA pixels
($\approx$19 au) in the blueshifted (approaching) direction relative the 
stellar trace in the 
unbinned  total-exposure G140M 2D spectral image.
Further discussion of the H$_{2}$ lines is given in Section 4.3.

\input{f1.tex}

\subsection{STIS G140M 1D Spectra Centered on the Star }

The STIS G140M 1D total-exposure spectrum extracted along the stellar trace 
is shown in Figure 2. The spectrum was extracted with a spatial binsize of
5 MAMA pixels (full-width = 0$''$.145) and thus captures emission out to offsets of
$\pm$0$''$.0725 ($\pm$9.7 au) from the star. 
The spectrum is dominated by the bright broad blueshifted 
C IV doublet lines (Table 3). But as noted above,  several 
slightly-blueshifted fluorescent H$_{2}$ lines are also detected (Table 4). 
In addition to the conspicuous H$_{2}$ lines visible in Figure 2,
the blue wing of  C IV 1548 \AA~ shows a narrow peak that is 
due to  H$_{2}$ R(3)1-8 emission and C IV 1551 \AA~ may contain
a weak contribution from H$_{2}$ P(8)2-8. Also visible in the
G140M spectrum is a feature at rest-frame wavelength 
1533.51 $\pm$ 0.04 \AA~ identified as Si II ($\lambda_{lab}$ = 1533.43 \AA~).
Fainter  emission from the Si II line at
$\lambda_{lab}$ = 1526.71 \AA~ may also be  present.

\input{f2.tex}

Figure 3 zooms in on the C IV doublet in the  G140M 1D on-source spectrum.
Both  lines are broad and noticeably asymmetric. Trial fits of 
each line with a simple 1-component Gaussian model begin to reproduce the 
overall broad shapes but residuals remain at the wavelengths expected for
H$_{2}$ R(3)1-8 and P(8)2-8. Adding narrow low-velocity Gaussians  
with velocities and widths fixed at  $-$9 km s$^{-1}$ and FWHM = 44 km s$^{-1}$
based on mean values of other H$_{2}$ lines in the spectrum (Table 4)
recovers the excess and results in a reasonably good first approximation of
the overall doublet shape (Fig. 3). 
The two H$_{2}$ lines that overlap C IV in the on-source spectrum contribute only 6\%  to
the observed (absorbed) flux in the wavelength range spanned by the C IV
doublet and most of this comes from the R(3) 1-8 line. 

\input{f3.tex}

\input{table3.tex}

Line widths of FWHM =  304 $\pm$ 26 km s$^{-1}$ are needed to reproduce the 
broad C IV lines, which is well in excess of that expected from stellar rotation at
$v$sin$i$ = 52 $\pm$ 2 km s$^{-1}$ (Petrov et al. 1999).
As a result of the 5 MAMA pixel spatial binning, the spectrum extracted
on the stellar trace also captures emission from the inner jet ($<$9.7 au) which 
undoubtedly contributes to the line broadening. The Gaussian-fit centroids of the 
C IV lines in the stellar trace spectrum  are  blueshifted to 
$-65$ $\pm$ 10 km s$^{-1}$. However, this value cannot be considered a
reliable flow velocity because of the asymmetric line shapes and blending.
In addition, the  flux ratio of the C IV doublet lines in the stellar trace spectrum
is significantly less than the value F$_{1548}$/F$_{1551}$ =  2
expected for optically thin emission (Table 3) so optical depth effects
may be important near the star. A similar conclusion was reached by Lamzin (2000).
More reliable velocities are obtained
from off-source spectra of the jet where the C IV lines are narrower and well-modeled
as single-component Gaussians (Sec. 3.4).

\input{table4.tex}

Even though there is little doubt that the spectrum extracted along
the stellar trace includes C IV emission from the inner jet near the star,
it is not obvious that the star itself contributes. There is 
no significant peak at the stellar radial velocity in the broad 
C IV line profile shown in Figure 3.
Nevertheless, estimates of the C IV flux contributed by the 
inner unresolved jet to the on-source spectrum are significantly less 
than the measured flux, so there is little doubt that the star itself
is contributing (Sec. 3.4).

We compared  the C IV doublet flux in 14 exposures with individual exposure times of 
10 - 14 minutes acquired over a timespan of 7.8 hours. No deviations of
greater than 1.9$\sigma$ with respect to the mean were found but low-level
flux variations of up to $\approx$20\% cannot be ruled out because of
the lower S/N ratio in spectra of individual exposures as compared 
to their sum. Since the C IV emission is extended on a spatial
scale of at least one arcsecond the measured flux depends on the
aperture size or the spatial bin width used in STIS spectral extractions
and comparisons with flux measurements
from previous studies are not straightforward. But we note that the
observed flux measured in a total exposure spectrum centered on the stellar
trace and using a spatial binsize of 2$''$ to capture the central source
and  essentially all of  the detected jet emission out to offsets of  $\pm$1$''$ is
F$_{CIV,absorbed}$(1548$+$1551) = 8.65 $\pm$ 0.58 $\times$ 10$^{-14}$ ergs cm$^{-2}$ s$^{-1}$
(H$_{2}$ subtracted).  For comparison, the value obtained by {\em IUE}
was F$_{CIV,absorbed}$ = 8.7 $\pm$ 0.3 $\times$ 10$^{-14}$ ergs cm$^{-2}$ s$^{-1}$
(Valenti et al. 2000). The agreement is remarkably
good considering that most of the {\em IUE} observations were obtained
in the 1980s.

The feature in Figure 2 identified as Si II deserves further comment.
This line was also listed as a detection in {\em HST}  Goddard High-Resolution
Spectrograph (GHRS) spectra of RY Tau analyzed by Lamzin (2000).
It forms at a characteristic temperature T$_{SiII}$ $\sim$ 10$^{4.5}$ K
and thus traces slightly cooler plasma than C IV. 
As can be seen in Figure 2 the line consists of a sharply-peaked Gaussian 
superimposed on a broad pedestal extending from $\approx$1532.5 - 1534.5 \AA~.
A Gaussian fit of the sharply-peaked core gives a redshifted 
velocity of $+$16 $\pm$ 6 km s$^{-1}$ and FWHM = 72 $\pm$ 12 km s$^{-1}$. 
The apparent redshift is surprising
since all other identified lines in the G140M on-source spectrum are
blueshifted. It is thus obvious that the line does not form in the
approaching jet and it may originate near the star in the receding
(redshifted) jet.

\subsection{Extinction and C IV Luminosity}
The intrinsic (unabsorbed) C IV line flux and luminosity  are quite uncertain
due to discrepancies in the value of $A_{\rm v}$  determined from
different color indices (Calvet et al. 2004) and subtleties in the
far-UV extinction law toward  dark clouds such as
Taurus (Fitzpatrick \& Massa 1988; Cardelli, Clayton, \& Mathis 1989; Whittet et al. 2004).
Previous UV studies of RY Tau have adopted different stellar extinction  values
including  $A_{\rm v}$ = 0.29 (Ardila et al. 2002), 0.55 (G\'{o}mez de Castro \& Verdugo 2007),
0.6 - 1.3  (Lamzin 2000), 1.0 - 1.3 (Petrov et al. 1999), 1.8 (Kenyon \& Hartmann 1995)
and 2.2 $\pm$ 0.2 (Calvet et al. 2004).
The latter value is clearly larger than the others but is nevertheless consistent
with independent estimates based on X-ray absorption (Skinner et al. 2016).
Assuming an extinction ratio $A_{\rm{1550~ \AA~}}$/$A_{\rm v}$ $\approx$ 2.6 (Calvet et al. 2004),
our absorbed fluxes convert to unabsorbed values according to
F$_{C IV,unabsorbed}$ = 10$^{1.04 A_{\rm v}}$~F$_{C IV,absorbed}$.
At d = 134 pc the intrinsic line luminosity is
L$_{C IV}$ = 2.146$\times$10$^{42}$~F$_{C IV,unabsorbed}$~ergs s$^{-1}$.
If the extinction is at the high end of the range of values in the literature ($A_{\rm v}$ $\approx$ 2)
then the absorbed line flux for the star$+$jet out to offsets of $\pm$1$''$ of
F$_{CIV,absorbed}$(1548$+$1551) = 8.65  $\times$ 10$^{-14}$ ergs cm$^{-2}$ s$^{-1}$
converts to an intrinsic (unabsorbed) luminosity  
L$_{C IV}$(star$+$jet) $\sim$ 2 $\times$10$^{31}$ ergs s$^{-1}$,
which is comparable to the  X-ray luminosity
of RY Tau determined from {\em Chandra} and {\em XMM-Newton}  observations (Table 1).
The above derivation assumes that $A_{\rm v}$ is the same for the star and the
jet out to a separation of 1$''$.

\input{f4.tex}

\input{f5.tex}

\input{f6.tex}

\subsection{Spectra Extracted Along the Jet}

Figure 4 shows the C IV doublet in  1D total-exposure spectra extracted at 
an offset of 0$''$.29  (10 MAMA pixels) from the star along the blue and redshifted jets
using a spatial bin width of  0$''$.145 (5  pixels). These 
spectra capture emission at offsets of 0$''$.29 $\pm$ 0$''$.0725 (38.9 $\pm$ 9.7 au)
along the jets.
The lines in the blueshifted jet provide higher S/N ratio and more reliable fit 
parameters. As is evident from the figure, the C IV lines are narrower and
more symmetric (nearly Gaussian) than in the on-source spectrum. There is very 
little H$_{2}$ contamination at this offset.  H$_{2}$ line fluxes are down by
more than  an order-of-magnitude compared to the on-source
spectrum and are negligible.
Gaussian fits of the C IV line at an offset of 0$''$.29 in
the blueshifted jet give a velocity that is 9\% larger than that of
the corresponding redshifted jet  but the blue and red jet velocities
agree to within the measurement uncertainties (Table 3).
Thus, we find no evidence for a 
significant velocity asymmetry in the jet at an offset of 0$''$.29 (39 au).
The ratio of observed fluxes in the 
doublet lines in the spectra extracted at this offset is consistent
with the value F$_{1548}$/F$_{1551}$ = 2  expected for optically thin emission.

Figure 5 plots the absorbed C IV flux (sum of the doublet components)
in the blueshifted jet as a function of projected offset from the star. 
The flux decreases toward larger offsets and the falloff is well-fitted
by a simple exponential model at offsets $\geq$0$''$29. 
The jet flux cannot be reliably measured at small offsets $<$0$''$29 or at 
offsets beyond 1$''$. However, extrapolating the exponential fit back
to the star gives a predicted blue-jet flux 
F$_{C IV,absorbed}^{blue-jet}$(offset=0$''$) =
1.51 $\times$ 10$^{-14}$ ergs cm$^{-2}$ s$^{-1}$.
We assume that the red-jet contributes equally to the  C IV flux
in the on-source  spectrum but this is probably an overestimate since
the red-jet flux is weaker than the blue-jet at offsets away from
the star (Table 3). We thus expect an on-source  C IV flux contribution
from the blue and red jet lobes combined of no more than twice the 
above value, namely 
F$_{C IV,absorbed}^{blue+red-jets}$(offset=0$''$0) =
3 $\times$ 10$^{-14}$ ergs cm$^{-2}$ s$^{-1}$.
But the absorbed flux measured in the on-source spectrum
extracted using a 5 MAMA pixel spatial bin width  is
F$_{C IV,absorbed}^{total}$(offset=0$''$) =
5.41 $\times$ 10$^{-14}$ ergs cm$^{-2}$ s$^{-1}$ (Table 3).
The additional C IV flux in the on-source spectrum above that
expected from the inner jet lobes is very likely due to the
star itself. This result is not surprising since C IV emission is almost 
always detected in cTTS, even those without 
jets (Valenti, Johns-Krull, \& Linsky 2000; Ardila et al. 2013).

Figure 6 shows the radial velocity in the blueshifted jet as a function
of offset from the star as measured by the centroid of the brighter  
CIV 1548.2 \AA~ line. There is a clear decrease in jet speed toward larger 
offsets and this decrease is well-approximated by a linear fit 
v$_{blue\_jet}$ = 71.46 ($\pm$21.20) $\times$ offset (arcsecs) $-$ 164.7 ($\pm$11.0) km s$^{-1}$
for offsets in the range 0.$''$29 - 0.$''$73 (39 - 98 au).

\newpage

\section{Discussion}

\subsection{Jet Mass-Loss Rate}

The jet mass-loss rate for a fully-ionized jet is
\begin{equation}
\dot{M}_{jet} = \mu m_{H}n_{e}v_{jet}A
\end{equation}
where $\mu$ $\approx$ 1.24 is the mean atomic weight (amu) per nucleus
for gas with cosmic abundances,
$m_{H}$ is the proton mass, $n_{e}$ is the average electron
density in the jet, $v_{jet}$ is the (deprojected) jet velocity,
and $A$ is the cross-sectional area of the jet.
For simplicity we consider the blue jet lobe within
1$''$ of the star where our velocity measurements are
reliable. Within the inner arcsecond the jet is well-collimated
with nearly-constant deconvolved width FWHM $\approx$ 0$''$.18
(Agra-Amboage et al. 2009; Coffey et al. 2015). We thus assume
a cylindrical jet with cross-sectional area $A$ = 1.02 $\times$ 10$^{29}$ cm$^2$
at a distance of 134 pc.
At an offset of 0$''$.5 from the star the radial 
jet velocity is  $\approx$ 129 km s$^{-1}$ (Fig. 6).
We assume a jet inclination relative to the line-of-sight
$i_{jet}$ = 61$^{\circ}$ $\pm$ 16$^{\circ}$
as determined by Agra-Amboage et al. (2009). But for comparison
we note that Isella, Carpenter, \& Sargent (2010) used thermal
dust models based on multiwavelength millimeter data to constrain
the disk rotation axis inclination to be in the range  
$i_{disk}$ = 66$^{\circ}$ $\pm$ 2$^{\circ}$ (1.3 mm) -
71$^{\circ}$ $\pm$ 6$^{\circ}$ (2.8 mm).
For our adopted jet inclination angle the deprojected jet velocity is
v$_{jet}$ $\approx$ 266$^{+307}_{-84}$  km s$^{-1}$ where the
uncertainty is due only to the uncertainty in $i_{jet}$
and the much smaller uncertainty ($<$10\%) in the jet velocity
measurement from the C IV centroid shift is ignored. 
Substituting these numbers into
eq. (1) gives
\begin{equation}
\dot{M}_{jet,blue} = 8.9 \times 10^{-10}\left[\frac{n_{e}}{10^4~ \rm{cm^{-3}}}\right]~~\rm{M_{\odot} yr^{-1}}~.
\end{equation}

To estimate $n_{e}$ we use the expression for the C IV volume emission measure (EM) 
\begin{equation}
EM_{CIV} = n_{e}^2fV = L_{CIV}\Lambda^{-1}_{CIV}
\end{equation}
where $V$ is the volume of the emitting region, $f$ is the volume
filling factor of the C IV plasma (0 $<$ $f$ $\leq$ 1), L$_{CIV}$ is the unabsorbed
C IV luminosity, and $\Lambda_{CIV}$ = 6.8 $\times$ 10$^{-23}$ ergs s$^{-1}$ cm$^{3}$
is the volumetric cooling rate for C IV determined  by Schneider et al. (2013a).
As we have noted (Sec. 3.3), the value of L$_{CIV}$ for RY Tau is
subject to uncertainties in $A_{\rm v}$. Assuming $A_{\rm v}$ $\approx$ 2
and using the integrated C IV line flux within the inner arcsecond
of the blue jet (Fig. 5) we obtain EM$_{CIV}$ = 1.57 $\times$ 10$^{53}$ cm$^{-3}$
and $n_{e}\sqrt{f}$ = 2.6 $\times$ 10$^{4}$ cm$^{-3}$. Leaving $f$ as a
free parameter gives
$\dot{M}_{jet,blue}$ = (2.3 $\times$ 10$^{-9}$/$\sqrt{f}$) M$_{\odot}$ yr$^{-1}$.
This value is consistent with the value
$\dot{M}_{jet,blue}$ = (1.6 - 26)  $\times$ 10$^{-9}$ M$_{\odot}$ yr$^{-1}$
obtained by Agra-Amboage et al. (2009) using high spatial resolution
[OI]~$\lambda$6300 \AA~ spectral-images. For a symmetric jet
the derived mass-loss rate for blue and red jet lobes combined
would be about twice the above value.
The main uncertainties in the above calculation are the
deprojected jet velocity v$_{jet}$ ($i_{jet}$-dependent),  $L_{CIV}$ ($A_{\rm v}$-dependent),
and the volume filling factor $f$. In addition, we have assumed a fully-ionized
jet and if that is not the case then eq. (1) would need to be modified as
discussed by Hartigan, Morse, \& Raymond (1994).
Considering the uncertainties, the above can only be regarded as an order-of-magnitude
estimate of the jet mass-loss rate. 
For comparison, the mass-loss rate of RY Tau estimated by Kuhi (1964) 
using a spherically-symmetric isotropic wind model was
$\dot{M}_{*}$ = 3.1 (1.3 - 5.0) $\times$ 10$^{-8}$ M$_{\odot}$ yr$^{-1}$
where the range in parentheses reflects values obtained using different
spectral lines. A more recent estimate of the wind mass-loss rate by
Petrov, Babina, \& Artemenko (2017) gives
$\dot{M}_{wind}$ = 2 $\times$ 10$^{-9}$ M$_{\odot}$ yr$^{-1}$.

\subsection{Jet Heating Models}

The physical process which heats the jet and where the
heating occurs are not yet well-understood. The STIS
observation  discussed above reveals warm jet plasma
from C IV emission to within a few tenths of
an arcsecond of RY Tau, so some jet heating must take place
close to the star.  The new STIS data along with 
information on jet morphology obtained in earlier
studies allow some constraints on heating mechanisms
to be obtained, as discussed below.

{\em Base-heated Jet}:~
If heat input is restricted to the jet base near the star 
then it is launched hot and  cools by different
processes such as radiation and expansion as it flows
outward. Since the width of the RY Tau jet remains nearly constant 
within  $\sim$150 au ($\sim$1$''$) of the star (Agra-Amboage et al. 2009) 
we assume that radiative cooling is the dominant process. 
Any additional adiabatic cooling due to expansion would 
shorten the cooling times derived below, which should therefore
be regarded as upper limits.

The radiative cooling time is
\begin{equation}
\tau_{rad} = \frac{{\rm 3kT}}{n_{e}\Lambda(T)} ~. 
\end{equation}
We first assume a jet-base temperature  
T$_{jet}$ $\sim$ T$_{CIV}$ $\sim$ 10$^{5}$ K
but  this should be considered a lower limit
and the possibility of a higher jet-base
temperature is considered below. For a solar-abundance plasma we use 
$\Lambda$(T = 10$^{5}$ K) = 6.5 $\times$ 10$^{-22}$ ergs s$^{-1}$ cm$^{3}$,
of which about one-third comes from carbon-cooling transitions,
primarily C III (977 \AA~) and C IV (Foster et al. 2012;
Lykins et al. 2013; Smith et al. 2001). Inserting this value
into eq. (4) along with 
n$_{e}$$\sqrt{f}$ = 2.6 $\times$ 10$^{4}$ cm$^{-3}$ (Sec. 4.1)
gives a radiative cooling time $\tau_{rad}$ = 0.077$\sqrt{f}$ yr.

The distance traversed by the jet during the cooling time depends 
on the (deprojected) jet velocity after taking into account the
jet inclination angle $i_{jet}$ relative to the line-of-sight.
The jet inclination is not well-determined for RY Tau but previous work 
suggests $i_{jet}$ = 61$^{\circ}$ $\pm$ 16$^{\circ}$ (Agra-Amboage et al. 2009).
At an offset of 0$''$.5 from the star, which we take as representative,
the radial velocity of the blue jet is 129 km s$^{-1}$ (Fig. 6).
This corresponds to a deprojected velocity  v$_{jet}$ = 266 (182 - 573) km s$^{-1}$
and a projected (tangential) velocity v$_{jet,t}$ = 233 (159 - 501) km s$^{-1}$,
where the range in parentheses reflects the uncertainty in $i_{jet}$.
At this speed, the time required for the jet to traverse a projected distance 
of 134 au (1$''$) is 2.7 (1.2 - 4.0) yr. But during the cooling time
the jet traverses a projected  distance of only  
$d_{t}$  = 3.8 (2.6 - 8.1)$\sqrt{f}$ au.
This value is much less than the distance $d_{t}$  $\sim$ 134 au (1$''$)
to which the blueshifted jet is actually detected in C IV (Fig. 1).

The above discrepancy could be resolved if the radiative 
cooling time  is longer than  estimated above, or if the jet 
undergoes additional heating as it flows outward. Possible additional  
heating  mechanisms are discussed  below.
A longer radiative cooling time could be achieved if
the jet base temperature  is significantly higher
than the  value T$_{jet}$ $\sim$ T$_{CIV}$ $\sim$ 10$^{5}$ K
assumed above, or if n$_{e}$ is lower than assumed.
Another question that may be relevant to the above discrepancy 
is whether the jet plasma is in collisional ionization equilibrium.

Some models assume temperatures at the jet-base of
a few MK as in the disk corona heating model 
of  Takasao, Suzuki, \& Shibata (2017). Their
analysis of the DG Tau jet adopted a jet-base
temperature T$_{jet}$ = 3.4 MK based on the X-ray
observations  of G\"{u}del et al. (2008).
The jet-base temperature of RY Tau is not known but
a  value of at least T$_{jet}$ $\sim$ 3 - 4 MK
(kT $\sim$ 0.26 - 0.34 keV) would be needed to produce 
detectable jet thermal X-ray emission above {\em Chandra's} 
low-energy limit. Such emission may be 
present at offsets out to $\approx$1$''$.7 from the star along
the blueshifted jet (Fig. 4 of Skinner et al. 2011).
However, {\em Chandra}'s angular resolution (FWHM $\approx$ 0$''$.5)
is insufficient to spatially distinguish between X-rays
originating near the jet base within a few au of RY Tau
or further out in the jet.

Assuming  a jet-base temperature T$_{jet}$ $\sim$ 3 MK
at the low-end of {\em Chandra's} detection range gives
a radiative cooling time (eq. [4]) of
$\tau_{rad}$ = 6.9$\sqrt{f_{x}}$/(n$_{e}$/10$^{5}$ cm$^{-3}$) yr.
Here $f_{x}$ is the volume filling factor of X-ray emitting
plasma in the jet and we have used  $\Lambda$= 
$\Lambda_{x}$(T = 3 MK) = 5.7 $\times$ 10$^{-23}$ ergs s$^{-1}$ cm$^{3}$ 
for solar abundance plasma (Foster et al. 2012). 
To produce soft X-ray
emission  out to a projected (tangential) separation of
1$''$.7 from the star, the cooling time would 
need to be at least $\tau_{rad}$ $\sim$ 4.6 yr for
v$_{jet}$ = 266 km s$^{-1}$. This would be the
case if n$_{e}$ $\ltsimeq$ 1.5 $\times$ 10$^{5}$$\sqrt{f_{x}}$ cm$^{-3}$. 
This electron density is consistent with that derived above
using the C IV luminosity provided that $f_{x}$ $\geq$ 0.03.

As is clear from the above, the ability of the base-heated jet 
model to produce soft X-rays at offsets $>$1$''$ for a given
jet speed depends critically on the jet plasma cooling time. 
The radiative cooling time is proportional to 
T$_{jet}$/n$_{e}$$\Lambda(T)$, but neither T$_{jet}$ nor
n$_{e}$ is tightly-constrained by observations.
For the above estimate we assumed T$_{jet}$ $\sim$ 3 MK 
but if T$_{jet}$ $\sim$ 6 MK then $\tau_{rad}$ increases 
by a factor of 2.5 for a given n$_{e}$. Non-radiative
cooling from processes such as expansion may also
contribute but has been neglected above since the optical
jet width shows little change at the projected offsets
of interest here and reliable measurements of the
width of the faint X-ray jet versus projected separation
are not available.

{\em Collisional Ionization Equilibrium}:~A difference between
the observed spatial extent of the C IV jet and that predicted from
the radiative cooling time $\tau_{rad}$ could occur if the jet plasma
is not in collisional ionization equilibrium (CIE). Departures
from CIE occur when plasma is rapidly heated (e.g. by shocks) or cooled 
on a timescale much shorter than the time $\tau_{cie}$ for ionization equilibrium 
to be restored. Under non-CIE conditions the plasma
is in a transient state during which it radiates at a temperature
that is decoupled from ionization balance (Mewe 1999).

The timescale $\tau_{\rm cie}$ required for a given element to 
reach CIE is inversely proportional to electron density $n_{e}$ and has a 
complex electron-temperature dependence arising from differences in
ionization and recombination rates of various ions 
(Fig. 1 of Smith \& Hughes 2010; Mewe 1999). Without accurate
estimates of the RY Tau jet temperature and electron density
we cannot reach a firm conclusion as to whether it is in CIE
but a preliminary assessment is possible based on 
order-of-magnitude estimates.

If the jet temperature is  T $\approx$ 10$^{5}$ K, 
the metals which would take the longest time to reach CIE are
C and Al (Smith \& Hughes 2010). At this temperature, the time 
required for all C ions to be within  90\% of their CIE value is
$\tau_{\rm cie,c}$ $\approx$ 3.8/(n$_{e}$/10$^{4}$ cm$^{-3}$) yr.
Using our estimate $n_{e}\sqrt{f}$ = 2.6 $\times$ 10$^{4}$ cm$^{-3}$ (Sec. 4.1)
gives $\tau_{\rm cie,c}$ $\approx$ 1.5$\sqrt{f}$ yr.
As noted previously, the time required for the jet to
propagate out to the C IV-traced distance of 134 au (1$''$) is 
$\tau_{jet}$ = 2.7 (1.2 - 4.0) yr. Thus, if the jet temperature is
T $\approx$ 10$^{5}$ K then CIE appears likely. 
If the jet is at X-ray emitting temperatures 
T $\gtsimeq$ 3 $\times$ 10$^{6}$ K then the value of $\tau_{\rm cie,c}$
determined by Smith \& Hughes (2010) is at least an order-of-magnitude
less than at 10$^{5}$ K and the case for CIE is strengthened.
Based on the above estimates, the case for invoking non-CIE
conditions in the RY Tau jet does not look compelling.

{\em Hot Plasmoids}:~
Ejection of hot plasmoids during powerful stellar X-ray 
flares  could produce high-temperature plasma in 
discrete structures  offset from the star. This picture
is reminiscent of solar coronal mass ejections and
has been discussed in the context of large X-ray flares
from protostars by Hayashi, Shibata, \& Matsumoto (1996).

In their picture discrete plasmoids at or near coronal
flare-loop temperatures are ejected at high velocities
of several hundred km s$^{-1}$ but in extreme cases may
approach $\sim$1000 km s$^{-1}$. We assume that the 
ejected plasmoid is at temperature 
T$_{plasmoid}$ $\sim$ 10$^{8}$ K since superhot 
X-ray flares at such temperatures have been
detected in RY Tau (Skinner et al. 2011).
The radiative cooling time from eq. (4) is
then $\tau_{rad}$ = 0.044/(n$_{e}$/10$^{9}$ cm$^{-3}$)~~yr
where for superhot solar abundance  X-ray plasma $\Lambda$(T = 100 MK) =
2.9 $\times$ 10$^{-23}$ ergs s$^{-1}$ cm$^{3}$ (Foster et al. 2012).
The actual cooling time could be less since other processes
such as expansion may also cool the plasmoid.

X-ray spectra of cTTS give coronal  densities in the range 
n$_{e}$ $\approx$ 10$^{9.5}$ -  10$^{12}$ cm$^{-3}$ (Ness et al. 2004).
Adopting n$_{e}$ $\approx$ 10$^{9}$ cm$^{-3}$ as a lower 
limit on plasmoid density gives 
$\tau_{rad}$ $\ltsimeq$ 0.044 yr ($\ltsimeq$ 16 days).
Even at very high ejection speeds of $\sim$1000 km s$^{-1}$
the plasmoid would traverse a distance of only $\sim$10 au
during the cooling time, or 0$''$.075 at the distance of 
RY Tau. As the plasmoids cool down they  could be revealed 
in high-resolution optical images as knots close to the star. 
Because of their anticipated intermittent appearance during 
large X-ray flares  and short cooling times, plasmoids do not offer a 
plausible explanation for the omnipresent jet of RY Tau
that is traced outward to $>$100 au.

{\em Shock-heated Jet}:~
The maximum shock temperature for a shock-heated
jet is $T_{s}$ = 0.15(v$_{s}$/100 km s$^{-1}$)$^2$ MK, where 
v$_{s}$ $<$ v$_{jet}$ is the shock 
speed (Raga, Noriega-Crespo, \& Vel\'{a}zquez 2002).
As shown in Figure 6 the projected  speed in the blueshifted 
jet decreases with increasing distance from the star.
For a representative offset of 0$''$.5,
the deprojected jet speed allowing for the uncertainty
in $i_{jet}$ is v$_{jet}$ = 266 (182 - 573) km s$^{-1}$.  
At these speeds the maximum shock temperature is
$T_{s}$ = 1.1 (0.5 - 4.9) MK.
These temperatures are high enough to account for 
C IV emission from shocks (T$_{CIV}$ $\sim$ 0.1 MK) but 
can only explain hotter X-ray  plasma T$_{x}$ $\sim$ few MK
if the jet speed is at the high end of the expected range
and v$_{s}$ $\approx$ v$_{jet}$. On the contrary,
Agra-Amboage et al. (2009) have argued that shock 
speeds are much less than the jet speed. 

At offsets  of $\geq$1$''$, where faint X-ray emission is 
visible in {\em Chandra} images, the jet will have decelerated to 
lower speeds than the value at 0$''$.5 used above and the 
maximum shock temperature will also be lower. Extrapolation of 
the linear decline shown in Figure 6 to an offset of 1$''$ results in a
jet speed that is about 32\% less than given above and a
corresponding maximum shock temperature T$_{s}$ $\approx$ 2.3 MK
(kT$_{s}$ $\approx$ 0.2 keV). This value is at or
below  {\em Chandra}'s low-energy threshold. Thus, the
production of soft X-rays at offsets $\geq$1$''$ in the
RY Tau jet within {\em Chandra}'s detectability range
is problematic if the jet velocity profile based on
C IV measurements shown in Figure 6 is taken as a reference.

{\em Magnetic Jet Heating}:~
Currents generated by wound-up magnetic fields could play a role
in heating ionized jets. 
But magnetic fields in young stellar jets
are expected to be weak ($\mu$G - mG) and hence difficult to detect
and measure (Hartigan et al. 2007).
Few good-quality measurements exist but an important result
is the detection of radio synchrotron radiation in the powerful 
Herbig-Haro jet HH 80-81 by Carrasco-Gonz\'{a}lez et al. (2010).
This provides clear evidence that the jet has a magnetic field.
Their radio data imply an equipartition magnetic field strength
in jet knots of $B_{jet}$ $\approx$ 0.2 mG. But knots trace 
overdense regions in the jet and theoretical models predict that 
$B_{jet}$ increases with density  (Hartigan et al. 2007). Thus, 
a non-uniform magnetic field that is stronger in dense knot
regions is expected and the spatially-averaged magnetic field
strength in HH 80-81 is probably less than the above jet-knot value.

To our knowledge, there has been no  B-field  measurement for the 
RY Tau jet and we  thus have to rely on estimates. If the jet
is magnetically-confined then the magnetic pressure must
exceed the gas (plasma) pressure, P$_{gas}$ = $\beta$P$_{mag}$
where 0 $<$ $\beta$ $<$ 1.
Specifically, 
$n_{tot}$kT$_{jet}$ = $\beta$$(B^2_{jet}$/8$\pi$) 
where $n_{tot}$ is the total  particle density in
the jet, k is Boltzmann's constant, and T$_{jet}$ is 
the jet temperature. We assume T$_{jet}$ $\sim$ T$_{CIV}$ $\sim$ 10$^{5}$ K
as above but as already noted this should be taken as a lower limit.
For solar-abundance plasma with fully-ionized H and He we get
$n_{tot}$ $\approx$ 1.92$n_{e}$ $\approx$ (5 $\times$ 10$^{4}$)/$\sqrt{f}$
when using the value of $n_{e}$$\sqrt{f}$ obtained in Sec. 4.1.
This gives  $B_{jet}$ = 4/($f^{0.25}$$\beta^{0.5}$) ~mG.
For marginal confinement ($\beta$ = 1), $B_{jet}$ = 4/$f^{0.25}$~mG
but strong-confinement values $\beta$ $\ll$ 1 are typically used in 
magnetic jet simulations.

To obtain a second estimate of $B_{jet}$ we use the canonical formula
for magnetized young stellar jets proposed by Hartigan et al. (2007),
\begin{equation}
\frac{B_{jet}}{15~ \mu G} = \left[\frac{n_{tot}}{100~ {\rm cm}^{-3}}\right]^{0.85}~.
\end{equation}
Using $n_{tot}$ $\approx$ (5 $\times$ 10$^{4}$)/$\sqrt{f}$ for RY Tau as 
above gives $B_{jet}$ $\approx$ 3/$f^{0.425}$~mG, similar to that obtained
using equipartition arguments.  

Adopting a deprojected jet speed v$_{jet}$ $\approx$ 266 km s$^{-1}$
at an offset of 0$''$.5 from the star as a representative value and a
jet radius $r_{jet}$ $\approx$ 0$''$.09 $\approx$ 12 au for the
inner arcsecond of the RY Tau jet, the magnetic energy in the 
jet is  $L_{mag}$ $\sim$ 10$^{30}$($B_{jet}$/5 mG)$^{2}$~ergs s$^{-1}$,
where we have used the results of Schneider et al. (2013a) but
have normalized to a weaker field strength of 5 mG. Using this
result the minimum required field strength to account for 
the integrated blue jet C IV luminosity (Fig. 5) is 
$B_{jet}$ $\sim$  11 mG if  A$_{\rm v}$ $\approx$ 2 
(i.e. L$_{CIV,blue-jet}$ = 1.1 $\times$ 10$^{31}$ ergs s$^{-1}$).
This field strength is about 50 times greater 
than determined  for the HH 80-81 jet by
Carrasco-Gonz\'{a}lez et al. (2010).

\subsection{Fluorescent H$_{2}$ Lines}

Fluorescent H$_{2}$ lines are commonly present in the FUV spectra of TTS
and are thought to be pumped mainly by wings of the stellar H Ly$\alpha$ line.
Detailed studies of fluorescent H$_{2}$ emission in selected samples of TTS
with the {\em HST} were carried out by Ardila et al. (2002) using the GHRS  and
France et al. (2012) using the Cosmic Origins Spectrograph (COS) and STIS. 
Their results revealed that the 
H$_{2}$ lines in cTTS are broadened to typical values 
FWHM $\approx$ 40 $\pm$ 20 km s$^{-1}$ with velocities ranging from
0 km s$^{-1}$ (i.e. the line centroid is centered at the stellar
radial velocity) to values of a few tens of km s$^{-1}$. 

The H$_{2}$ lines that we detect  for RY Tau are blueshifted
by an average of  $-$9 $\pm$ 1 km s$^{-1}$ and moderately broadened
to an average width  FWHM = 44 $\pm$ 2 km s$^{-1}$. The brightest
H$_{2}$ line is R(11) 2-8 as shown in Figure 7. 
The velocity shifts are significantly larger than the STIS G140M wavelength
calibration uncertainty (Table 2). The line broadening is also significant
compared to the STIS G140M instrumental resolution of FWHM $\approx$ 15 km s$^{-1}$
at 1500 \AA. The fitted H$_{2}$ line-widths are well above that expected 
from thermal broadening  and  turbulence, as has generally been found for the larger 
samples of TTS analyzed  in the studies referenced above.  Thus, bulk motion of 
molecular gas is implied.

\input{f7.tex}

Several different origins of the fluorescent H$_{2}$ lines in cTTS
have been proposed. In the sample analyzed by Ardila et al. (2002)
the H$_{2}$ line centroids in most stars were blueshifted and line formation
in outflows was considered to be the most likely explanation. In the sample 
studied by France et al. (2012) some cTTS showed  H$_{2}$ lines with  
non-zero velocities suggestive of outflows, a good example being
the jet-driving cTTS RW Aur A. On the other hand, no significant velocity 
shift was detected in some cTTS and it was proposed that their fluorescent 
H$_{2}$ emission originates in a rotating disk. A detailed study of the jet-driving
cTTS DG Tau using STIS G140M (Schneider et al. 2013a) and {\em HST}
FUV imaging (Schneider et al. 2013b) revealed H$_{2}$ emission close
to the star near the approaching jet axis as well as  extended emission 
fanning out into a wide-angle conical-shaped structure. The wide-angle 
emission was attributed to a disk wind. In view of the various results 
and interpretations above, it seems likely that different regions near the star 
can contribute to fluorescent H$_{2}$ emission. The complex H$_{2}$ morphology of
DG Tau seen in {\em HST} FUV images supports this.

The remarkable similarity in blueshifted velocities and widths of the H$_{2}$
lines we detect in RY Tau, along with the faint blueward spatial extension seen
in brighter H$_{2}$ lines (Sec. 3.1), is indicative of a low-velocity
molecular outflow approaching the observer. The blueshifted  centroids 
are at odds with the broadened unshifted lines expected for H$_{2}$ 
formation in a rotating disk. The spatial morphology of
the outflowing gas is not well-determined by the narrow 0$''$.2 STIS
slit but H$_{2}$ emission is certainly present within 0$''$.0725 (9.7 au) of the star
(Fig. 2). In addition, the R(11) 2-8 and P(5) 1-8 lines are faintly visible
in G140M spectra  extracted at a blueward offset along the jet of 0$''$.29 $\pm$0$''$.0725
(38.9 $\pm$ 9.7 au) so the presence of H$_{2}$ out to a projected distance
of at least $\approx$30 au from the star is assured. The extended blueshifted 
H$_{2}$ emission may originate in molecular gas that is being swept outward
by the jet, possibly similar to the H$_{2}$ emission detected near the 
approaching jet axis of DG Tau (Fig. 3 of Schneider et al. 2013b). High-resolution
FUV images of RY Tau are needed to further clarify the spatial structure 
of the H$_{2}$ emission, especially away from the jet-axis where
extended conical-shaped structure has been seen in DG Tau.

\section{Summary}

The {\em HST} STIS observations presented here provide 
new information on RY Tau and its jet that help constrain
jet-heating models. The main results of this study are
summarized as follows.

\begin{enumerate} 

\item ~Warm  plasma (T$_{C IV}$ $\sim$ 10$^{5}$ K)
traced by C IV emission is clearly  detected in STIS 2D images out to a projected  offset
of 1$''$ (134 au) from  the star along the blueshifted  jet and out
to 0$''$.5 (67 au) along the fainter
redshifted jet. The C IV emission is traced inward to offsets of a
few tenths of an arcsecond, implying that  significant
jet heating occurs close to the star.

\item~The radial velocity in the brighter blueshifted jet
is $-$136 $\pm$ 10 km s$^{-1}$ at a projected offset of 0$''$.29 (39 au)
and the jet speed decreases toward larger offsets. 
No significant difference in the radial velocity of the blue
and redshifted jets is seen at an offset of 0$''$.29.

\item The mass-loss rate in the blueshifted jet is
$\dot{M}_{jet,blue}$ = 2.3 $\times$ 10$^{-9}$/$\sqrt{f}$~ M$_{\odot}$ yr$^{-1}$,
where $f$ is the volume filling factor of C IV plasma in the jet.
This value is consistent with previous optical measurements.

\item Several  fluorescent H$_{2}$ 
lines  blueshifted to $-$9 $\pm$ 1 km s$^{-1}$ are detected 
near the star. These lines evidently originate in  a low-velocity molecular
outflow but additional high spatial resolution FUV images are
needed to further clarify the outflow geometry.

\item Jet velocities determined from C IV centroids are
high enough to account for warm C IV plasma by shocks
in the inner jet at offsets of $<$1$''$. Soft
X-rays due to  shocks in  the inner jet
could also occur if the deprojected jet speed is near
the high end of the range allowed by jet inclination uncertainties
but extreme shock speeds v$_{s}$ $\approx$ v$_{jet}$  are required. 
At offsets beyond 1$''$ the extrapolated jet speed
is too low to produce soft X-rays at energies
above {\em Chandra}'s threshold.
Thus, shocks alone do not provide a satisfactory 
explanation of the soft X-ray emission detected 
by {\em Chandra} at offsets $>$1$''$ along the 
blueshifted jet.

\item Other jet-heating mechanisms besides shocks 
seem necessary to produce soft X-ray emission along
the jet. One possibility is that the jet is heated
to X-ray temperatures  near the base and is launched hot.
The heating process is unspecified but could well involve
magnetic processes in a magnetically-active star like
RY Tau. A recent study focusing on the DG Tau jet has 
proposed that a hot magnetic
disk corona could provide the base heating (Takasao et al. 2017).
Tighter observational constraints on the temperature at the jet base,
the jet electron density, and the  X-ray jet morphology are 
needed to assess the relevance of this picture to RY Tau. 
Even if the jet is launched at high temperatures, additional
heating may occur downstream. A jet magnetic field 
of minimum strength B$_{jet}$ $\sim$ 11 mG could account
for the jet C IV luminosity but the detection of such
weak fields is challenging.

\end{enumerate}

\acknowledgments

This work was supported by {\em HST} award HST-GO-13714 issued by the 
Space Telescope Science Institute (STScI) and is based on observations
made with the NASA/ESA Hubble Space Telescope, operated by 
the Association of Universities for Research in Astronomy, Inc. under 
contract with NASA. This work has made use of data
analysis products including STSDAS and PyRAF produced by the STScI.

\vspace{5mm}
\facilities{HST(STIS)}

\clearpage


\end{document}

%% file: table1.tex
\begin{deluxetable}{lcccccccc}
\tabletypesize{\scriptsize}
\tablewidth{0pt}
\tablecaption{Properties of RY Tau}
\tablehead{
           \colhead{Sp. Type}               &
           \colhead{M$_{*}$}            &
           \colhead{V}            &
           \colhead{A$_{\rm V}$}            &
           \colhead{L$_{bol}$}            &
           \colhead{log L$_{x}$}            &
           \colhead{$v_{rad}$}     &
           \colhead{log $\dot{M}_{acc}$}            &
           \colhead{d}             \\
           \colhead{}                   &
           \colhead{(M$_{\odot}$)}         &
           \colhead{(mag)}                   &
           \colhead{(mag)}                   &
           \colhead{(L$_{\odot}$)}           &
           \colhead{(ergs s$^{-1}$)}           &
           \colhead{(km s$^{-1}$)}         &
           \colhead{(M$_{\odot}$ yr$^{-1}$)}     &
           \colhead{(pc)}              
}
\startdata
 F8 III - G1-2 IV  & 1.7 - 2.0   &  9.3 - 11 (v) & 2.2$\pm$0.2    & 15.3  & 30.5 - 31.2 & 18$\pm$1   & $-$7.3$\pm$0.3  & 134     \\
\enddata
\tablecomments{Data are from Agra-Amboage et al. (2009), Calvet et al. (2004), Kenyon \& Hartmann (1995),
Petrov et al. (1999), and Schegerer et al. (2008). 
Spectral type is uncertain (Holtzman, Herbst, \& Booth 1986). The visual extinction (A$_{\rm V}$)
is from Calvet et al. (2004) but smaller values have been used in other studies (see text).
The bolometric luminosity is based on  L$_{bol}$ = 16.7 L$_{\odot}$ (d = 140 pc) from
Kenyon \& Hartmann (1995) and has been normalized to d = 134 pc. 
The X-ray luminosity L$_{x}$ is from Skinner, Audard, \& G\"{u}del (2016) and is variable.
The heliocentric radial velocity ($v_{rad}$) is from Petrov et al. (1999) and is consistent with an earlier
measurement $v_{rad}$ = 16.5 $\pm$ 2.4 km s$^{-1}$ obtained by Hartmann et al. (1986). 
Distance is from Bertout et al. (1999). }
\end{deluxetable}

%% file: table2.tex
\begin{deluxetable}{lccccccc}
\tabletypesize{\scriptsize}
\tablecaption{STIS G140M  Grating Observatations of RY Tau}
\tablehead{
           \colhead{Start}               &
           \colhead{Stop}               &
           \colhead{$\lambda$}      &
           \colhead{Resolution}            &
           \colhead{Plate scale}            &
           \colhead{Dispersion}     &
           \colhead{Exposures}            &
           \colhead{Time}             \\
           \colhead{(UT)}         &
           \colhead{(UT)}                   &
           \colhead{(\AA)}           &
           \colhead{FWHM (\AA)}           &
           \colhead{($''$/pixel)}         &
           \colhead{(\AA/pixel)}     &
           \colhead{ } &
           \colhead{(s)}              
}
\startdata
 2014-12-04 23:53  & 2014-12-05 07:39  & 1522-1578\tablenotemark{a}  & 0.075 & 0.029  & 0.05 & 14\tablenotemark{b}   &  10,765\tablenotemark{b}     \\
\enddata
\tablecomments{The observations were centered on RY Tau using target coordinates (J2000)
   R.A. = 04$^{h}$ 21$^{m}$ 57.41$^{s}$, Decl. = $+$28$^{\circ}$ 26$'$ 35.57$''$.
   All exposures were obtained in ACCUM mode with the 52$''$ x 0.2$''$ slit aligned along the optical
   jet axis (PA\_APER = 112.45$^{\circ}$). STIS target acquisition centers the source in the slit to
   an accuracy of  0.$''$01 (2$\sigma$). No additional peakup was performed.
   A 3-step STIS\_ALONG\_SLIT dither pattern was used
   with a 0.275$''$ step size. The spectral resolution for MAMA first-order modes is 1.5 pixels (FWHM) at 1500~\AA,
   equivalent to 15 km s$^{-1}$ (FWHM).  
   The STIS G140M  wavelength accuracy from calibration studies  expressed as the mean difference between observed 
   (Gaussian-fitted) and laboratory wavelengths of selected emission lines is 0.13 $\pm$ 0.14 (1$\sigma$) 
   MAMA pixels (Sonnentrucker 2015).  At 1500 \AA~ this equates to 1.3 $\pm$ 1.4 km s$^{-1}$.
   Instrument properties are from the {\em STIS Data Handbook} unless otherwise noted.
}
\tablenotetext{a}{For tilt corresponding to central wavelength $\lambda_{c}$ = 1550~\AA.}
\tablenotetext{b}{Exposure OCKF01060 (815 s) was not usable and is excluded from totals.}
\end{deluxetable}

%% file: f1.tex
\begin{figure}
\figurenum{1}
\epsscale{1.0}
\hspace{1.6in}
\includegraphics*[width=10.0cm,height=7.19cm,angle=0]{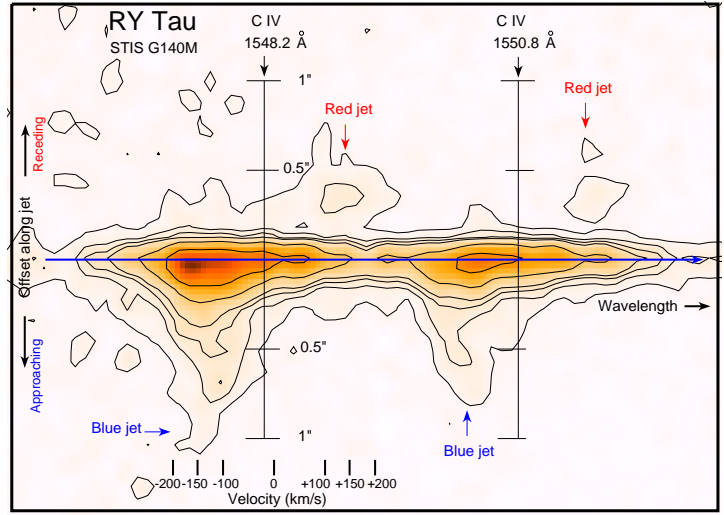} \\
\caption{
STIS MAMA G140M Gaussian-smoothed 2D spatially-resolved spectral-image  of the RY Tau jet
created by summing individual  exposures showing extended C IV (1548.2/1550.8~\AA~) emission.
The horizontal axis is wavelength (\AA~) and the vertical axis is spatial offset
from star along the jet (arcsecs). The horizontal line marks the position of the
stellar trace to within an uncertainty of $\pm$0.2 MAMA pixels ($\pm$0$''$.0058).
For the 1548.2 \AA~ component, the zero-velocity  tick mark corresponds to
1548.29 \AA~ after correcting for the star's heliocentric radial velocity ($v_{rad}$ = $+$18 km s$^{-1}$).
STIS spatial resolution is $\approx$0$''$.05 per 2-pixel spatial resolution element.
Contour levels are (2,4,6,8,16,32) $\times$ 10$^{-13}$ ergs cm$^{-2}$ s$^{-1}$~\AA$^{-1}$ arcsec$^{-2}$.
The lowest level contour is $\approx$2$\sigma$.
}
\end{figure}

%% file: f2.tex
\begin{figure}
\figurenum{2}
\epsscale{1.0}
\hspace{1.2in}
\includegraphics*[width=8.88cm,height=12.0cm,angle=-90]{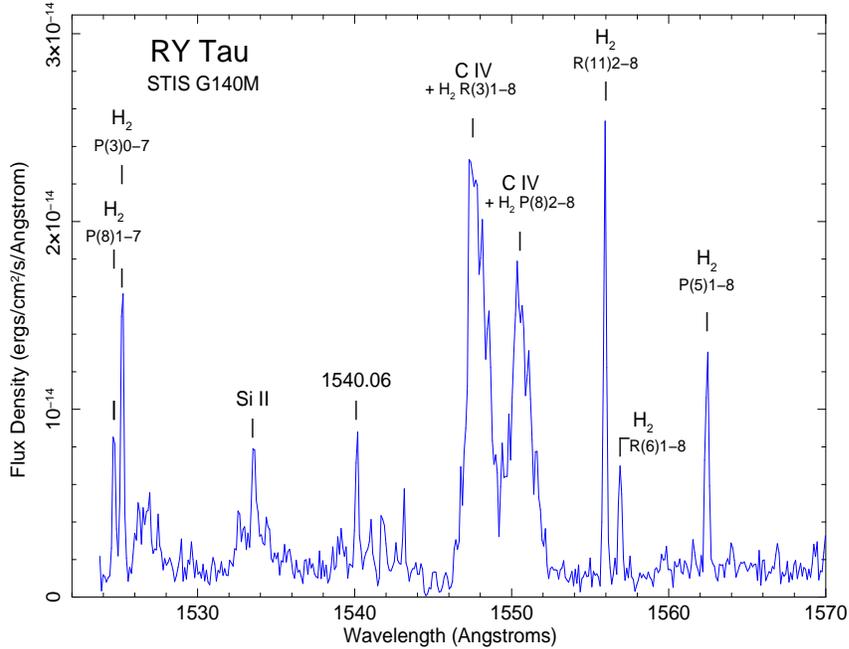} \\
\caption{STIS G140M RY Tau 1D spectrum showing flux density versus observed wavelength created from
all  usable exposures (10,765 s). The spectrum  was extracted
along  the stellar trace with a spatial bin full-width of 5 MAMA pixels (0$''$.145), capturing emission
out to offsets of  $\pm$ 0$''$.0725 = $\pm$9.7 au from the star.
The spectrum has been rebinned in wavelength by a factor of two for display. Error bars are omitted for
clarity. The unidentified  feature at rest-frame wavelength 1540.06 \AA~ may be  H$_{2}$ R(2)3-9.
}
\end{figure}

%% file: f3.tex
\begin{figure}
\figurenum{3}
\epsscale{1.0}
\hspace{1.5in}
\includegraphics*[width=7.4cm,height=10.0cm,angle=-90]{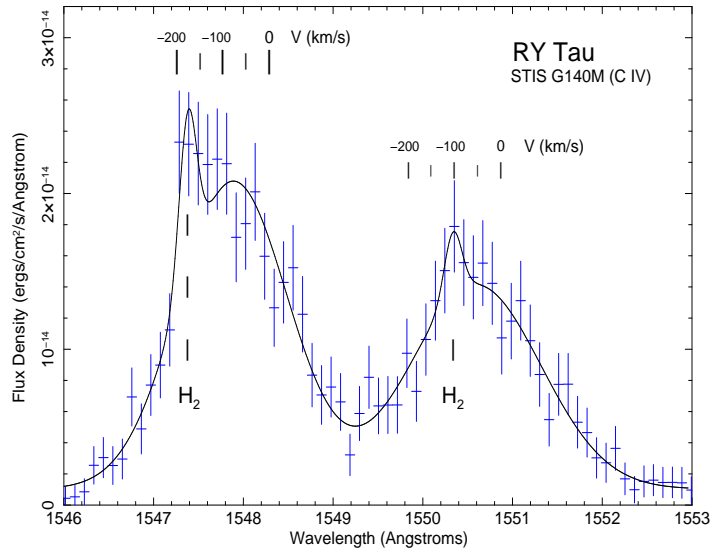} \\
\caption{
A zoomed version of Figure 2 showing  the C IV lines overlaid with a 4-component
Gaussian fit. Two components model the C IV doublet emission and two components
model the small contribution from the narrow fluorescent H$_{2}$ lines.
The velocity scales refer to the C IV lines and zero-velocity
wavelengths in the star's rest frame correspond to 1548.29 \AA~ and 1550.87 \AA~.
The velocities and line widths of the (R3)1-8 and P(8)2-8 fluorescent
H$_{2}$ lines were held fixed at $-$9 km s$^{-1}$ and FWHM = 44 km s$^{-1}$,
based on the mean values of other H$_{2}$ lines in the G140M spectrum (Table 4).
The C IV lines require Gaussian widths FWHM = 304 $\pm$ 26 km s$^{-1}$.
The absorbed flux of the C IV doublet (minus the H$_{2}$ contribution ) is
F(CIV)$_{\rm 1548+1551}$ = 5.41e-14 ergs cm$^{-2}$ s$^{-1}$.
The absorbed flux of the two H$_{2}$ lines combined is
F(H$_{2}$)$_{\rm 1547.3+1550.3}$ = 3.5e-15 ergs cm$^{-2}$ s$^{-1}$,
of which 71\% comes from the 1547.34 \AA~ R(3)1-8 line.
}
\end{figure}

%% file: table3.tex
\begin{deluxetable}{lccccc}
\tabletypesize{\scriptsize}
\tablecaption{RY Tau C IV Line Properties (STIS G140M)}
\tablehead{
           \colhead{Region}   &
           \colhead{Offset\tablenotemark{a}}               &
           \colhead{Velocity\tablenotemark{b,c}}                 &
           \colhead{FWHM\tablenotemark{c}}                     &
           \colhead{Total Flux\tablenotemark{d}}           &                      
           \colhead{Flux ratio\tablenotemark{e}}             \\
           \colhead{} &
           \colhead{(arcsecs) [au]}                         &
           \colhead{(km s$^{-1}$)}                   &
           \colhead{(km s$^{-1}$) }           &
           \colhead{(10$^{-14}$ ergs cm$^{-2}$ s$^{-1}$)}            &
           \colhead{}    
}
\startdata
star$+$jet      & ~~~0 [0]       & $-$65 $\pm$ 10\tablenotemark{f}   & 304 $\pm$ 26\tablenotemark{f}    & 5.41 $\pm$ 0.36\tablenotemark{g}  & 1.2 $\pm$ 0.2  \\
blue jet        &$-$0.29 [38.9] & $-$136 $\pm$ 10 & 127 $\pm$ 14    & 0.56 $\pm$ 0.05     & 2.5 $\pm$ 1.0 \\
red jet         &$+$0.29 [38.9] & $+$125 $\pm$ 14 & 127 $\pm$ 24    & 0.18 $\pm$ 0.02     & 2.0 $\pm$ 1.2 \\
\enddata
\tablenotetext{a}{Offset of the central pixel of the spectral extraction relative
to the stellar trace. A positive offset is in the direction of the redshifted (receding)
jet. The spectra were extracted using a spatial binsize of 5 MAMA pixels (0$''$.145 full width) and
capture emission within  $\pm$0$''$.0725 of the extraction center. Thus, the spectrum
centered on the stellar trace (offset=0) captures emission from both the  blue and red
inner jet lobes.}
\tablenotetext{b}{Centroid velocity is in the star's rest frame and assumes
a heliocentric stellar radial velocity of  $+$18 km s$^{-1}$. }
\tablenotetext{c}{Average of the two lines comprising the doublet.}
\tablenotetext{d}{Sum of the observed (absorbed) continuum-subtracted flux of the two lines in the doublet.
                  The H$_{2}$ flux contribution has also been subtracted for the spectrum centered on
                  the stellar trace. At offsets of $\pm$0$''$.29 the H$_{2}$ flux is negligible (Sec. 3.4).}
\tablenotetext{e}{Ratio of the observed fluxes of the doublet lines:  F$_{1548}$/F$_{1551}$.}
\tablenotetext{f}{Lines are asymmetric (non-Gaussian) so the velocity  measurement
may not correspond to a physical flow speed. }
\tablenotetext{g}{A spectrum extracted using a larger spatial bin full-width of 69 MAMA pixels (2.$''$0)
captures the star and jet emission out to offsets of $\pm$1$''$ and gives a  total flux (H$_{2}$-subtracted)
F$_{CIV,absorbed}$(1548$+$1551) = 8.65 $\pm$ 0.58 $\times$ 10$^{-14}$ ergs cm$^{-2}$ s$^{-1}$.   }
\end{deluxetable}

%% file: table4.tex
\begin{deluxetable}{lcccc}
\tabletypesize{\scriptsize}
\tablecaption{RY Tau H$_{2}$ Lines (STIS G140M)}
\tablehead{
           \colhead{Transition\tablenotemark{a}}               &
           \colhead{$\lambda_{lab}$}                           &
           \colhead{Velocity\tablenotemark{b}}                 &
           \colhead{FWHM\tablenotemark{c}}                     &
           \colhead{Flux\tablenotemark{d}} \\                     
           \colhead{}                         &
           \colhead{(\AA~)}                   &
           \colhead{(km s$^{-1}$) }           &
           \colhead{(km s$^{-1}$)}            &
           \colhead{(10$^{-15}$ergs cm$^{-2}$ s$^{-1}$)}    
}
\startdata
P(8) 1-7  & 1524.65 & $-$10$\pm$2    & 42$\pm$2     & 1.9 $\pm$ 0.7  \\
P(3) 0-7  & 1525.15 & $-$8$\pm$2     & 42$\pm$2     & 3.8 $\pm$ 0.8  \\
R(3) 1-8  & 1547.34 & [$-$9]\tablenotemark{e}        & [44]\tablenotemark{e}    & 2.5 $\pm$ 1.5  \\
P(8) 2-8  & 1550.30 & [$-$9]\tablenotemark{e}        & [44]\tablenotemark{e}    & 1.0 $\pm$ 0.6  \\
R(11) 2-8 & 1555.89 & $-$8$\pm$2     & 42$\pm$3     & 5.1 $\pm$ 0.9  \\
R(6) 1-8  & 1556.87 & $-$10$\pm$2    & 45$\pm$4     & 1.5 $\pm$ 0.7  \\
P(5) 1-8\tablenotemark{f}  & 1562.39 & $-$7$\pm$2     & 47$\pm$3      & 3.4 $\pm$ 0.9  \\ 
mean (s.d.)\tablenotemark{g} & ...   & $-$9 (1)       & 44 (2)       & ....        \\
\enddata
\tablenotetext{a}{Transitions are in the Lyman band system (Abgrall et al. 1993).
The letter denotes the change in rotational quantum number ($\Delta$J) from that in the
upper electronic state (J$'$) to the lower electronic state (J$''$) in the sense
$\Delta$J = J$'$ $-$ J$''$. The letter P corresponds to $\Delta$J = $-$1 and
R corresponds to $\Delta$J = $+$1. The number in parentheses following
the letter is the rotational quantum number in the lower electronic state (J$''$). The two
numbers separated by a hyphen correspond to the vibrational quantum numbers 
$v'$-$v''$ in the  upper and lower electronic states, respectively.
}
\tablenotetext{b}{Centroid velocity is in the star's rest frame and assumes
a heliocentric stellar radial velocity of  $+$18 km s$^{-1}$. }
\tablenotetext{c}{STIS G140M  resolution is FWHM = 15 km s$^{-1}$.}
\tablenotetext{d}{Observed (absorbed) continuum-subtracted flux;  not corrected for extinction.}
\tablenotetext{e}{Quantities in brackets were held fixed during fitting. 
                  Blended with C IV.}
\tablenotetext{f}{This line has been detected in some Herbig-Haro objects such as HH 43/47 (Schwartz 1983).}
\tablenotetext{g}{The blended R(3) 1-8 and P(8) 2-8 lines were excluded from calculation of the mean.}
\end{deluxetable}

%% file: f4.tex
\begin{figure}
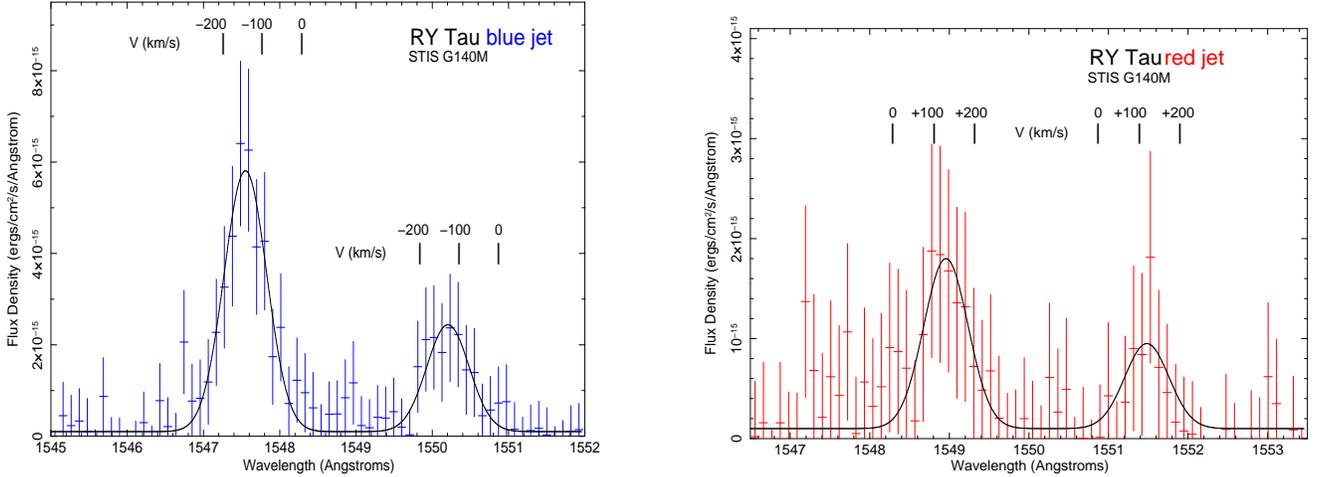

\figurenum{4}
\epsscale{1.0}
\includegraphics*[width=6.29cm,height=8.5cm,angle=-90]{f4l.eps}
\hspace{0.5cm} 
\includegraphics*[width=6.29cm,height=8.5cm,angle=-90]{f4r.eps}
\caption{G140M C IV total-exposure spectra of the RY Tau jet  showing flux density versus observed
wavelength.  The spectra were extracted  at an
offset of 10 MAMA pixels using a spatial bin full-width of 5 pixels,  capturing jet
emission at offsets of  0$''$.29 $\pm$ 0$''$.0725 (38.9 $\pm$ 9.7 au) from the star.
The spectra have been rebinned by a factor of two in wavelength.
Zero velocities correspond to 1548.29 \AA~ and 1550.87 \AA~ after correcting for the
star's heliocentric radial velocity ($v_{rad}$ = $+$18 km s$^{-1}$).
~{\em Left}:~Blueshifted (approaching) jet.  Line centroids give velocity shifts of
$-$143 $\pm$ 7 km s$^{-1}$ (1548 \AA~ component) and $-$129 $\pm$ 13 km s$^{-1}$ (1551 \AA~ component).
Line widths are FWHM = 127 km s$^{-1}$. The observed (absorbed) integrated line fluxes are
F$_{1548}$ = 4.00e-15 and F$_{1551}$ = 1.58e-15 ergs cm$^{-2}$ s$^{-1}$.~
{\em Right}:~Redshifted (receding) jet. Line centroids give velocity shifts of
   $+$130 $\pm$ 15 km s$^{-1}$ (1548 \AA~ component) and $+$120 $\pm$ 13 km s$^{-1}$
   (1551 \AA~ component). Line widths are FWHM = 127 km s$^{-1}$.
The observed (absorbed) integrated line fluxes are
F$_{1548}$ = 1.18e-15 and F$_{1551}$ = 0.53e-15 ergs cm$^{-2}$ s$^{-1}$.~
}
\end{figure}

%% file: f5.tex
\begin{figure}
\figurenum{5}
\epsscale{1.0}
\hspace{1.3in}
\includegraphics*[width=7.4cm,height=10.0cm,angle=-90]{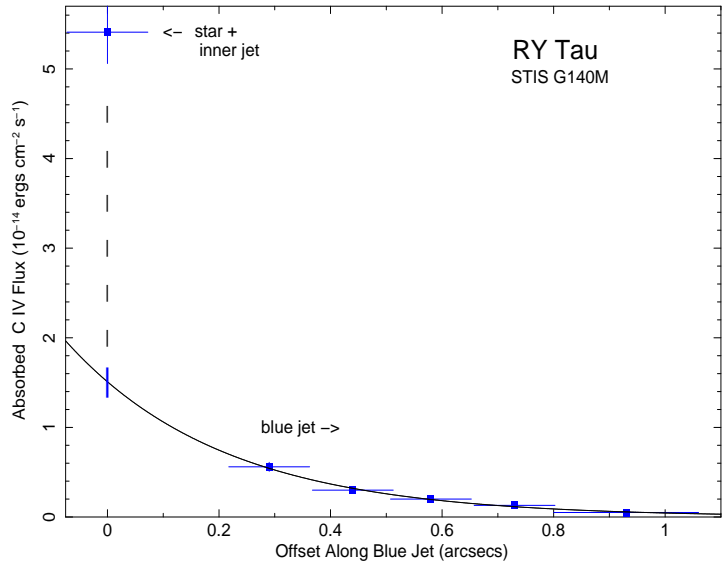}
\caption{
Absorbed C IV flux (sum of the 1548 \AA~ and 1551 \AA~ components) versus projected
offset from RY Tau along the blueshifted jet. Fluxes were measured in spectra extracted using a spatial
bin width of 5 MAMA pixels (full-width = 0$''$.145) except for the last data point
centered at offset = 0$''$.93 which was extracted using a bin width of 9 pixels
(full-width = 0$''$.26) to bring out faint jet emission. The data point for the
spectrum extracted on the stellar trace (offset = 0$''$) contains emission from the
star and the unresolved inner blue and redshifted jets out to offsets of
$\pm$0$''$.0725 ($\pm$9.7 au). The solid line is an exponential fit of
the five blue jet flux data points and gives
Absorbed CIV Flux = Aexp[$-$(offset $-$ B)/C] where the offset (arcsecs) is taken as a positive
value and best-fit parameters are
A = 0.1977, B = 0.5768, and C = 0.2838.
Integrating the exponential model over the range 0$''$ - 1$''$ gives a
total blue-jet flux F$_{C IV}$$^{blue-jet}$ = 4.15 $\times$ 10$^{-14}$ ergs cm$^{-2}$ s$^{-1}$.
Extrapolating the model back to the star  gives a blue jet flux at the zero offset position of
1.51 $\times$ 10$^{-14}$ ergs cm$^{-2}$ s$^{-1}$.
}
\end{figure}

%% file: f6.tex
\begin{figure}
\figurenum{6}
\epsscale{1.0}
\hspace{1.3in}
\includegraphics*[width=7.4cm,height=10.0cm,angle=-90]{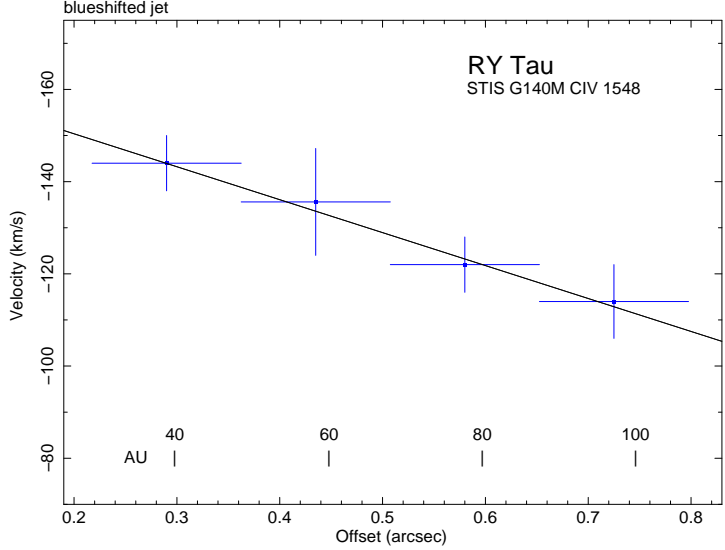}
\caption{
Radial velocity of the blueshifted jet versus projected offset from the star
along the jet. Velocities are based on centroid measurements of the C IV 1548.2 \AA~ line
in G140M spectra extracted at increasing offsets of 10, 15, 20, 25 MAMA pixels
(38.9, 58.3, 77.7, 97.1 au) along the jet and a spatial bin full-width of 5 pixels (0$''$.145).
Velocities  are relative to the star's rest frame assuming a stellar radial velocity
of $+$18 km s$^{-1}$.
The projected offset is converted to au assuming a distance of 134 pc.
The solid line is a linear fit of the four data points and gives
V = 71.46 ($\pm$21.20) $\times$ offset(arcsec) $-$ 164.7 ($\pm$11.0) km s$^{-1}$.
}
\end{figure}

%% file: f7.tex
\begin{figure}
\figurenum{7}
\hspace{1.3in}
\epsscale{1.0}
\includegraphics*[width=7.4cm,height=10.0cm,angle=-90]{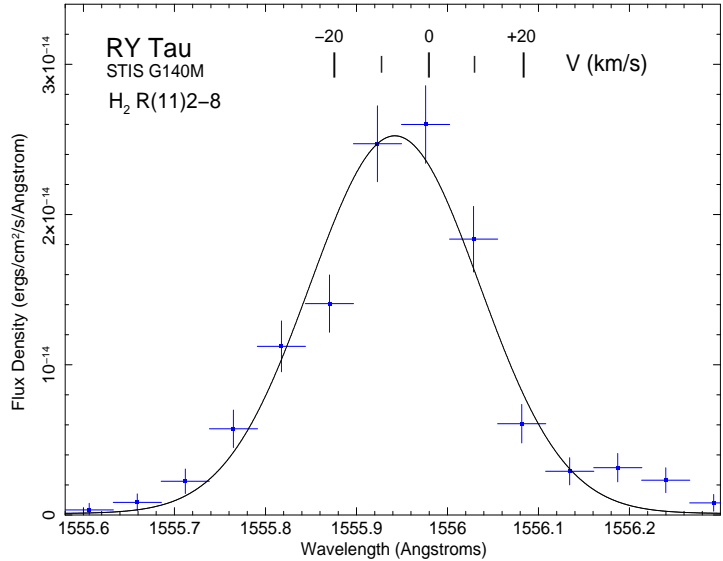} \\
\caption{
The RY Tau  H$_{2}$ R(11)2-8 line at full STIS G140M spectral
resolution (i.e. no rebinning in wavelength) from a spectrum extracted along
the stellar trace using a spatial bin width of 5 MAMA pixels (full-width = 0$''$.145).
The spatial width captures emission out to offsets of $\pm$0$''$.0725 ($\pm$9.7 au)
from the star along the jet. The velocity scale is relative to the
star's rest frame and zero-velocity corresponds to 1555.98 \AA~ for a heliocentric stellar
radial velocity of $+$18 km s$^{-1}$. The Gaussian fit gives a velocity of
$-$7.7 $\pm$ 2 km s$^{-1}$ and FWHM = 42 $\pm$ 3 km s$^{-1}$.
}
\end{figure}